\documentclass[%
 aip,
 amsmath,amssymb,
 reprint,%
]{revtex4-1}

\usepackage{graphicx}
\usepackage{dcolumn}
\usepackage{bm}

\usepackage[utf8]{inputenc}
\usepackage[T1]{fontenc}
\usepackage{mathptmx}
\usepackage{etoolbox}

\newcommand{\mean}[1]{\left\langle #1\right\rangle}

\makeatletter
\def\@email#1#2{%
 \endgroup
 \patchcmd{\titleblock@produce}
  {\frontmatter@RRAPformat}
  {\frontmatter@RRAPformat{\produce@RRAP{*#1\href{mailto:#2}{#2}}}\frontmatter@RRAPformat}
  {}{}
}%
\makeatother
\begin{document}

\preprint{AIP/123-QED}

\title[]{Fluctuations, correlations, and Casimir-like forces in the homogeneous cooling state of a granular gas}
\author{Jesús David Jiménez Oliva}
\affiliation{Universidad Complutense Madrid, 28040 Madrid, Spain}
\author{Pablo Rodriguez-Lopez}
\affiliation{\'Area de Electromagnetismo \& GISC, Universidad Rey Juan Carlos, 28933 M\'ostoles, Spain}
\author{Nagi Khalil}
\affiliation{\'Area de Electromagnetismo \& GISC, Universidad Rey Juan Carlos, 28933 M\'ostoles, Spain}
\email{\texttt{nagi.khalil@urjc.es}}


\date{\today}

\begin{abstract}
  The fluctuating hydrodynamics by Brey et. al. is analytically solved to get the long-time limit of the fluctuations of the number density, velocity field, and energy density around the homogeneous cooling state of a granular gas, under physical conditions where it keeps stable. Explicit expressions are given for the non-white contributions in the elastic limit. For small dissipation, the latter is shown to be much smaller than the inelastic contributions, in general. The fluctuation-induced Casimir-like forces on the walls of the system are calculated assuming a fluctuating pressure tensor resulting from perturbing its Navier-Stokes expression. This way, the Casimir-like forces emerges as the correlation between the longitudinal velocity and the energy density. Interestingly, the fluctuation-induced forces push/pull the system towards the square or rectangular geometry when they vanish, in good agreement with the event-driven numerical simulations. 
\end{abstract}

\maketitle

\section{\label{sec:level1}Introduction}

The Casimir effect is the appearance of quantum-thermal fluctuation-induced interactions of the electromagnetic field between objects. Casimir showed in his seminar works \cite{casimir1948attraction,casimir1948influence} that quantum fluctuations of the electromagnetic field, even at zero temperature, make two parallel plates in vacuum to attract each other. In this case, boundary conditions at the plates restrict the fluctuation modes, giving rise to an energy density in between the plates different form that outside and, importantly, dependent on the distance between the plates, which ultimately induces the attraction. The Casimir effect has been studied in different configurations, taking into account the properties of the involded materials \cite{Lifshitz1957} and the particular geometry of the experiment \cite{Casimir_Scattering2007,Scattering_formalism_em,Lambrecht2006}. See \cite{woods2016materials,dantchev2023fluctuation} for recent reviews.

However, the Casimir effect is not a purely electromagnetic interaction, but an emergent interaction of the electromagnetic field due to its fluctuations \cite{rodriguez2011dynamical}. 
Moreover, Casimir-like forces or fluctuation-induced forces can appear between objects immersed in any kind of field under fluctuations. For this reason, the Casimir effect can be understood as an universal interaction.

Fluctuation-induced interactions have been studied in many different physical systems \cite{kardar1999friction,gambassi2009casimir,woods2016materials} showing that two main ingredients can be identified for Casimir-like forces to arise: large enough fluctuations and a confining geometry \cite{PabloRL2011}. When we come to thermal, classical systems, the former condition typically requires being close to a critical point \cite{fisher1978wall}. This is the case of Casimir-like forces measured within a critical binary mixture \cite{krech1999fluctuation,hertlein2008direct,fisher2003phenomenes,gambassi2009critical}, in wetting films \cite{squarcini2023derivation,squarcini2022casimir}, Ising systems \cite{dantchev2023critical}, and Brownian motion of colloids \cite{maciolek2018collective}, just to mention a few examples. Moreover, there is a growing interest in the study of Casimir-like forces in out-of-equilibrium systems, given that many of them show non-negligible fluctuations even from critical points\cite{golestanian2005fluctuation}. Examples include Soret-Casimir effect for system driven by a thermal gradient \cite{najafi2004forces,kirkpatrick2013giant,kirkpatrick2014fluctuation}, liquid mixtures \cite{kirkpatrick2015nonequilibrium,kirkpatrick2016physical,kirkpatrick2016nonequilibrium}, reaction-diffusion systems \cite{brito2007generalized}, and active matter \cite{Gambassi2022}, among many others. 

Of particular relevance for our study is the granular Casimir effect postulated by Brito et. al. in \cite{Brito_Casimir_Granular,brito2007casimir} and further generalized to non-equilibrium systems \cite{rodriguez2011dynamical,rodriguez2011stochastic,PhDThesis:Rodriguez-Lopez-2011}. Brito and co-workers measured and calculated the Casimir-like force between two large intruders in a thermostated granular gas. Interestingly, according to their theory, the force between intruders originates form an unbalance renormalized pressure, due to density fluctuations and density-temperature correlations. Moreover, the existence of fluctuation-induced forces is not the approach to a critical point but an intrinsic property of the granular dynamics.

In this article, we study Casimir-like forces in the homogeneous cooling state (HCS) of a granular gas \cite{brey1996homogeneous,garzo1999homogeneous,goldhirsch2003homogeneous}. This state can be reached by a granular gas in a system with periodic boundary conditions, and is macroscopically characterized by being spatially homogeneous, with zero velocity field, and decreasing in time energy. The HCS is known to be fundamental to understand the macroscopic dynamics of a granular gases, by playing the role of a reference state around which the hydrodynamic description can be constructed \cite{brey1998hydrodynamics,khalil2014hydrodynamic,khalil2020unified}, for instance. Hence, advancing in the study of the fluctuation properties of the HCS seems natural. In particular, we consider a hypercube system and compute the fluctuation-induced forces acting on the "virtual" walls under physical conditions where the HCS keeps stable. This provides us with information about the role of fluctuations in the stability of the HCS. 

The rest of the paper is organized as follows. Section \ref{sec:fh} is devoted to the definitions and to introduce the fluctuating hydrodynamic description of the system in terms of the fluctuations of the number density, the velocity field, and the energy density. The values of most coefficients can be found in Appendix \ref{appen:1}.  We use the theory by Brey et. al. \cite{brey2009fluctuating,brey2011fluctuating}, systematically derived from the fluctuating Boltzmann equation for smooth inelastic hard spheres or disks. It has two main differences if compared to previous theories \cite{van1997mesoscopic,van1998spatial,van1999randomly}: it includes the contribution of the density gradient to the heat flux, which has been shown to be relevant under some physical conditions \cite{brey2004heat,brey2005heat,khalil2016heat}, and it also accounts for the non-white properties of the fluctuation terms as well as the breakdown of fluctuation-dissipation relations \cite{brey2008breakdown}. 

The fluctuating hydrodynamics is analytically solved in Sec.~\ref{sec:sc} to obtain the long-time limit of the fluctuations and correlations of the fluctuations of the hydrodynamic fields in the Fourier space. Explicit expressions are given for the elastic case with small wave vector, providing the first order contribution of the non-white properties of the sources, as well as the leading inelastic contributions for small dissipation and small wave vector. The long-time limit is considered because only in this case the amplitudes of the the fluctuations of the sources have an explicit known dependence on time, which in turn is sufficient for the subsequent computation of the Casimir-like forces. Further calculation details are given in Appendix \ref{appen:2}.

The results of Sec.~\ref{sec:sc} are used to derive an expression for the fluctuation-induced force on the "virtual" walls of the system in Sec.~\ref{sec:casimir}. We first propose an expression for the pressure tensor up to second order in the fluctuations and then derive the Casimir-like force using a microscopic cutoff, given by the particles diameter. Explicit expression of the force is given for the rectangular (2D) system, under some approximations as done in Appendix \ref{appen:3}. It is shown that the fluctuating contribution to the force on both the horizontal and vertical walls vanishes for a square system, being different from zero in other cases. The results are compared against event-driven molecular dynamics simulations. Details of the numerical simulations can be found in Appendix \ref{appen:4}. We finish with a discussion and some conclusions in Sec.~\ref{sec:conclusion}.

\section{\label{sec:fh}Fluctuating hydrodynamics}

The granular gas is modeled as \(N\) smooth hard spheres (\(d=3\)) or disks (\(d=2\)) in a rectangular container with volume \(V=L_1\cdots L_d\) and periodic boundary conditions. All particles have the same mass \(m\) and diameter \(\sigma\), moving freely between inelastic collisions characterized by a constant coefficient of normal restitution \(\alpha\in[0,1]\). The case \(\alpha=1\) corresponds to the elastic limit.

\subsection{Homogeneous cooling state}

It is well known that under periodic boundary conditions, when the system is small enough, a granular gas can reach the so-called homogeneous cooling state (HCS). From a macroscopic point of view, the HCS is characterized by a spatially homogeneous number density \(n_H=\frac{N}{V}\), zero mean velocity, and a decreasing-in-time granular temperature \(T_H(t)\). The latter is related with the total energy density \(\frac{d}{2}n_HT_H\) and obeys the Haff's law \cite{haff1983grain,khalil2018generalized}:
\begin{equation}
\partial_sT_H=-\zeta_0 T_H.  
\end{equation}
The time scale \(s\) is proportional to the number of accumulated collisions, defined as 
\begin{equation}
  ds= \frac{v_0}{\lambda}dt,
\end{equation}
where 
\begin{equation}
  v_0=\sqrt{\frac{2T_H}{m}}
\end{equation}
and 
\begin{equation}
\lambda= \frac{1}{n_H\sigma^{d-1}}
\end{equation}
are proportional to the thermal velocity and the mean free path, respectively. The dimensionless cooling rate \(\zeta_0\) is a function of \(\alpha\), whose approximate expression is given in Appendix \ref{appen:1}, and takes into account the dissipation of energy due the inelastic collisions. Hence, it vanishes in the elastic limit \(\alpha\to 1\).

\subsection{Fluctuations around the HCS}

As in the elastic case \cite{landau2013fluid}, the hydrodynamics quantities fluctuate around their values in the HCS. In \cite{brey2009fluctuating,brey2011fluctuating} closed fluctuating Navier-Stokes equations were derived for the fluctuations of the number density \(\delta n\), momentum density (proportional to the velocity field) \(\delta \boldsymbol {G}\), and the energy density \(\delta E\). These quantities depend on position \(\boldsymbol r\) and time \(s\).

Before providing the fluctuating equations, it is convenient to introduce dimensionless hydrodynamics quantities as
\begin{eqnarray}
  && \delta \rho(\boldsymbol \ell,s)=\frac{\delta n(\boldsymbol \ell,s)}{n_H}, \\
  && \delta \boldsymbol\omega(\boldsymbol \ell,s)=\frac{\delta G (\boldsymbol \ell,s) }{mn_Hv_0(s)}, \\
  && \delta \epsilon(\boldsymbol \ell,s)=\frac{\delta E(\boldsymbol \ell,s)}{\frac{d}{2}n_HT_H(s)},
\end{eqnarray}
where \(\boldsymbol \ell\) is a dimensionless vector measuring positions in unit of the local mean free path
\begin{equation}
  \boldsymbol \ell=\frac{\boldsymbol r}{\lambda}.
\end{equation}
The corresponding Fourier quantities, denoted by \(\delta \rho(\boldsymbol k,s)\), \(\delta \boldsymbol \omega(\boldsymbol k,s)\), and \(\delta \epsilon(\boldsymbol k,s)\) are given by
\begin{equation}
  \delta \rho(\boldsymbol \ell ,s)=\frac{1}{\mathcal V}\sum_{\boldsymbol k}\delta \rho({\boldsymbol k},s) e^{i\boldsymbol k\cdot \boldsymbol{\ell}},
\end{equation}
and similarly for the other two fluctuation fields. We have introduced the dimensionless volume 
\begin{equation}
\mathcal V=\mathcal L_1\dots \mathcal L_d,
\end{equation} 
with
\begin{equation}
  \mathcal L_i=\frac{L_i}{\lambda}=n_H\sigma^{d-1}L_i, \qquad i=1,\dots,d,
\end{equation}
the dimensionless lengths. Since we are assuming a finite system with periodic boundary conditions, the wave vectors of the Fourier sum are restricted such as 
\begin{equation}
  k_i=\frac{2\pi}{\mathcal L_i}n_i, \quad n_i\in \mathbb Z, \qquad i=1,\dots,d,
\end{equation}
Finally, the fluctuations are real magnitudes, hence
\begin{equation}
  \label{eq:realrho}
  [\delta \rho(\boldsymbol k,s)]^*=\delta \rho(-\boldsymbol k,s),
\end{equation}
and similarly for the velocity and energy fluctuations. 

\subsection{Fluctuating Navier-Stokes equations}
From the results in \cite{brey2011fluctuating}, the Langevin Navier-Stokes equations for the fluctuations of the hydrodynamic quantities in the Fourier space read
\begin{eqnarray}
  \label{eq:ns1}
  \partial_s\delta \rho&+&i\boldsymbol k\cdot \delta \boldsymbol \omega=0, \\
  \label{eq:ns2}
  \partial_s\delta \boldsymbol \omega &+&\left( \eta k^2-\frac{\zeta_0}{2}\right) \delta \boldsymbol \omega+ \eta \frac{d-2}{d}{\boldsymbol{k}} \boldsymbol{k}\cdot \delta \boldsymbol \omega  \nonumber \\ &+&i\frac{\boldsymbol k}{2}\delta \epsilon={\boldsymbol W},\\
  \label{eq:ns3}
  \partial_s\delta\epsilon&+&\left[\frac{\zeta_0}{2}-\frac{2}{d}k^2( \kappa- \mu)\right]\delta \rho+i\frac{d+2}{d}\boldsymbol k\cdot \delta \boldsymbol \omega \nonumber \\ &+&\left(\frac{\zeta_0}{2}+\frac{2}{d}k^2 \kappa\right)\delta \epsilon={\mathcal E},
\end{eqnarray}
where we have removed the dependence of the quantities on \(\boldsymbol k\) and \(s\) to easy the notation. The new quantities are the dimensionless transport coefficients, namely the viscosity \(\eta\), the thermal conductivity \(\kappa\), and the diffusive heat conductivity \(\mu\). They are time dependent quantities whose asymptotic values for large times are given in the Appendix \ref{appen:1}. Moreover, the two magnitudes on the right-hand side of the second and third equations are uncorrelated noise sources whose statistical properties read 
\begin{eqnarray}
  &&\mean{{\boldsymbol W}}=\boldsymbol 0, \\
  \label{eq:corrw}
  &&\mean{ W_i(\boldsymbol k,s) W_j(\boldsymbol k',s')}\simeq \frac{d}{2(d-1)} c_{w}\delta_{\boldsymbol k,-\boldsymbol k'}\nonumber \\ && \mbox{}\hspace{2cm} \times \left(\delta_{ij}+\frac{d-2}{d}\frac{k_ik_j}{k^2}\right)e^{\lambda_4|s-s'|}, \\
  &&\mean{{\mathcal E}}=0, \\
  \label{eq:correp}
  &&\mean{{\mathcal E}(\boldsymbol k,s){\mathcal E}(\boldsymbol k',s')}\simeq \left[c_{\epsilon}^{(1)}\delta(s-s') \right.\nonumber \\ && \mbox{}\hspace{3cm}\left.+c_{\epsilon}^{(2)}k^2 e^{\lambda_5|s-s'|}\right]\delta_{\boldsymbol k,-\boldsymbol k'}, 
\end{eqnarray}
where the approximate relations hold for \(s\gg 1,\, s'\gg 1\) and the new coefficients  \(c_w,\, c_\epsilon^{(1)}\), and \(c_\epsilon^{(2)}\) are explicitly given in Appendix \ref{appen:1}. The exponents \(\lambda_4\) and \(\lambda_5\) can be written as a function of the transport coefficients:
\begin{eqnarray}
\label{eq:lam4}
  &&\lambda_4\simeq -\frac{1}{2\eta}+\frac{\zeta_0}{2}, \\
\label{eq:lam5}
  &&\lambda_5\simeq -\frac{3(d+2)-2\kappa\zeta_0}{4(3\kappa-2\mu)}.
\end{eqnarray}

Note that, on the one hand, the fluctuating Navier-Stokes equations \eqref{eq:ns1}--\eqref{eq:ns3} are restricted to small gradients, i.e. small values of \(k\), as well as to small hydrodynamic fluctuations. Moreover, noise terms are no longer white but include a finite (exponential) correlation time. On the other hand, they include the elastic case for \(\alpha=1\). However, the Landau theory \cite{landau2013fluid} for the molecular fluids is recovered not only by taking \(\alpha=1\), but also by assuming that the noise terms are delta-correlated, which is true provided time correlations among the noise terms decay faster than the typical hydrodynamic time scales, namely
\begin{eqnarray}
  && e^{\lambda_4 s}\to \frac{2}{|\lambda_4|}\delta(s), \\
  && e^{\lambda_5 s} \to \frac{2}{|\lambda_5|}\delta(s). 
\end{eqnarray}
This is equivalent to assuming that \(|\lambda_4|s\gg 1, \,|\lambda_4|s\gg 1\) for \(s\) in the hydrodynamic time scale. 

\section{\label{sec:sc} Spatial correlations}

In this section we solve the fluctuating Navier-Stokes equations to get the steady-state spatial correlations of the fluctuating hydrodynamic quantities. In doing so, we first consider the general solution to the set \eqref{eq:ns1}--\eqref{eq:ns3}. Then we use it to get the structure factors.

\subsection{General solution }

In order to solve the fluctuating Navier-Stokes equations, it is useful to decompose the fluctuation of the velocity into its component along \(\boldsymbol k\) and normal to it:
\begin{eqnarray}
  && \delta \phi=-i\frac{\boldsymbol k}{k}\cdot \delta \boldsymbol \omega, \\
  && \delta \boldsymbol \varphi=-i\delta \boldsymbol \omega -\delta \phi \frac{\boldsymbol k}{k},
\end{eqnarray}
where the imaginary unit \(i\) has been introduced to make the coefficients of the resulting equations real.

By means of the new quantities, the equation for \(\delta \boldsymbol \varphi\) decouples form the rest:
\begin{equation}
  \label{eq:sns0}
  \partial_s\delta \boldsymbol \varphi+\left( \eta k^2-\frac{\zeta_0}{2}\right) \delta \boldsymbol \varphi={\boldsymbol W}_\perp,
\end{equation}
where
\begin{equation}
  {\boldsymbol W}_\perp=-i{\boldsymbol W}-W_\parallel \frac{\boldsymbol k}{k},
\end{equation}
and
\begin{equation}
  W_\parallel=-i{\boldsymbol W}\cdot \frac{\boldsymbol k}{k}.
\end{equation}
As for the reminder fluctuating quantities, we have a closed set of equations:
\begin{eqnarray}
  \label{eq:sns1}
  && \partial_s\delta \rho-k \delta \phi=0, \\
  \label{eq:sns2}
  &&\partial_s\delta \phi +\left[\frac{2(d-1)}{d} \eta k^2-\frac{\zeta_0}{2}\right]\delta \phi+\frac{k}{2}\delta \epsilon={W}_\parallel,\\
  \label{eq:sns3}
  && \partial_s\delta\epsilon+\left[\frac{\zeta_0}{2}-\frac{2}{d}(\kappa- \mu)k^2\right]\delta \rho-\frac{d+2}{d}k\delta \phi \nonumber \\ && \mbox{}\hspace{2.8cm} +\left(\frac{\zeta_0}{2}+\frac{2}{d}\kappa k^2 \right)\delta \epsilon={\mathcal E}.
\end{eqnarray}

The Langevin equation \eqref{eq:sns0} for the perpendicular component of the fluctuations of the velocity is decoupled from the rest and can be directly solved:
\begin{equation}
  \delta \boldsymbol \varphi(s,k)=\int_0^sds_1\,e^{-\left(\eta k^2-\frac{\zeta_0}{2}\right)(s-s_1)}\boldsymbol W_\perp(s_1,k),
\end{equation}
where we have assumed that \(\delta \boldsymbol \varphi(0)=\boldsymbol 0\). It is readily seen that \(\delta \boldsymbol \varphi\) diverges as time rises for wave vectors \(\boldsymbol k\) such as \(\eta k^2-\frac{\zeta_0}{2}<0\). Hence, in order for the linear regime described by system \eqref{eq:sns0}, \eqref{eq:sns1}--\eqref{eq:sns3} to hold for large times \(s\gg 1\), we have to restrict ourselves to wave vectors larger than \(k_m\):
\begin{equation}
\label{eq:km}
  k_m=\sqrt{\frac{\zeta_0}{2\eta}}.
\end{equation}
This restriction is related to the linear stability of the HCS, extensively discussed in the literature \cite{brey1998hydrodynamics,brey2008shear}. Note that the wave-vector limitation is irrelevant in the elastic limit, since \(\zeta_0\to 0\) for \(\alpha\to 1\), while \(\eta\) keeps positive. 

Proceeding analogously, the general solution to \eqref{eq:sns1}--\eqref{eq:sns3} can be written as
\begin{equation}
  \label{eq:deltau}
  \delta \boldsymbol u(s,k)=\int_0^s ds_1\, e^{(s-s_1)A}\boldsymbol b(s_1,k),
\end{equation} 
where we have introduced the following matrix notation
\begin{eqnarray}
  && \boldsymbol u=(\delta \rho,\delta \phi,\delta \epsilon)^t, \\
  \label{eq:matA}
  && A=
     \begin{pmatrix}
       0&k&0 \\
       0&-\nu&-\frac{k}{2} \\
       \beta&\frac{d+2}{d}k&-\gamma
     \end{pmatrix}
                               ,\\
  \label{eq:vecb}
  && \boldsymbol b=(0,W_\parallel,{\mathcal E})^t.
\end{eqnarray}
and 
\begin{eqnarray}
  && \nu=\frac{2(d-1)}{d}\eta k^2-\frac{\zeta_0}{2}, \\
  && \beta=\frac{2}{d}(\kappa-\mu)k^2-\frac{\zeta_0}{2}, \\
  && \gamma=\frac{2}{d}\kappa k^2+\frac{\zeta_0}{2}.
\end{eqnarray}
The previous coefficients depend on \(\alpha\) and \(k\).

It can be seen that the eigenvalues of \(A\) have all negative real parts for a wide range of values of \(\alpha\) (including \(\alpha=1\)) provided \(k>k_m\). Hence, under the latter conditions, Eq.~\eqref{eq:deltau} provides the time dependence of the fluctuations for all times.

\subsection{Structure factor}

From the previous results we directly obtain that \(\delta \boldsymbol \varphi\) is uncorrelated from the other fluctuating magnitudes. Using \(\mean{\cdot}\) to denote average over the noise around the HCS, we directly obtain that
\begin{equation}
  \mean{\delta \boldsymbol \varphi(s,k) \delta \boldsymbol u(s',k')}=0,
\end{equation}
since the noise terms \(\boldsymbol W_\perp, \, W_\parallel\), and \(\mathcal E\) are uncorrelated. On the other hand, after some calculations we obtain the two-time correlation function for \(\delta \boldsymbol \varphi\)
\begin{eqnarray}
  \mean{\delta \boldsymbol \varphi_i(s,k)\delta \boldsymbol \varphi_j(s',k')}&=&\frac{\mathcal V^2\eta}{2N\left(\eta k^2-\frac{\zeta_0}{2}\right)}k^2 \delta_{\boldsymbol k,-\boldsymbol k} \nonumber \\  && \quad \times \delta_{ij} e^{-\left(\eta k^2-\frac{\zeta_0}{2}\right)|s-s'|},
\end{eqnarray}
which holds for \(s,s'\gg 1\). We refer to \cite{brey2008breakdown,brey2009fluctuating} for a deeper analysis of the correlations of the transverse velocity.

Focusing on the correlations among the other fields, the components of \(\delta \boldsymbol u\), it is also readily seen that only components with opposite wave vectors can be correlated. Again, this is a direct consequence of the statistical properties of the nose terms given in Eqs.~\eqref{eq:corrw} and \eqref{eq:correp}. Hence, in order to account for the relevant correlations taking place at the same time, we consider the following structure-factor vector:
\begin{eqnarray}
  \label{eq:sfv}
  && \boldsymbol S=(\mean{\delta \rho\delta \rho^*},\mean{\delta \rho\delta \phi^*},\mean{\delta \rho\delta \epsilon^*}, \nonumber \\ && \qquad \mean{\delta \phi\delta \phi^*},\mean{\delta \phi\delta \epsilon^*},\mean{\delta \epsilon\delta \epsilon^*})^t,
\end{eqnarray}
where all quantities are functions of the time \(s\) and the wave vector \(\boldsymbol k\). In Appendix \ref{appen:2} we show that the steady-state value of \(\boldsymbol S\) is given by
\begin{equation}
  \label{eq:smm1q}
  \boldsymbol S=M^{-1}\boldsymbol Q,
\end{equation}
where
\begin{equation}
  \label{eq:matm}
  M=
  \begin{pmatrix}
    0&2k&0&0&0&0\\
    0&-\nu&-\frac{k}{2}&k&0&0\\
    \beta&\frac{d+2}{d}k&-\gamma&0&k&0\\
    0&0&0&-2\nu&-k&0\\
    0&\beta&0&\frac{d+2}{d}k&-\nu-\gamma&-\frac{k}{2}\\
    0&0&2\beta&0&\frac{2(d+2)}{d}k&-2\gamma
  \end{pmatrix}
\end{equation}    
and
\begin{eqnarray}
\label{eq:matq}
\boldsymbol{Q} = \left(\begin{array}{c}
0 \\
\frac{c_w(\gamma-\lambda_4)}{|\lambda_4I+A|}k\\
-\frac{c_{\epsilon}^{(2)}}{2|\lambda_5I+A|}k^2\\
-\frac{2c_w\lambda_4(\gamma-\lambda_4)}{|\lambda_4I+A|}\\
\frac{c_w(\beta-\frac{d+2}{d}\lambda_4)}{|\lambda_4I+A|}k+\frac{c_{\epsilon}^{(2)}\lambda_5}{2|\lambda_5I+A|}k\\
-c_\epsilon^{(1)}-\frac{2c_\epsilon^{(2)}\lambda_5(\nu-\lambda_5)}{|\lambda_5I+A|}
\end{array}
\right) \rightarrow \left(\begin{array}{c}
0\\
0\\
0\\
-\frac{2c_w}{|\lambda_4|}\\
0 \\
-c_\epsilon^{(1)}-\frac{2c_\epsilon^{(2)}}{|\lambda_5|}
\end{array}
\right).
\end{eqnarray}
The limiting expression for \(\boldsymbol Q\) holds for \(|\lambda_4|\) and \(|\lambda_5|\) much larger than the entries of \(A\).

\subsection{Correlations and fluctuations for \(k=0\)}
For \(k=0\) the matrix \(M\) is singular, and we have to solve Eq.~\eqref{eq:msq} directly, without inverting \(M\). It is not difficult to see that all correlations and fluctuations vanishes for \(k=0\) but the energy fluctuations
\begin{equation}
\mean{\delta \epsilon\delta \epsilon^*}_{k=0}=\frac{4\mathcal V^2}{N}a_{33}(\alpha).
\end{equation}
This expression provide the global energy fluctuation around the HCS, as already reported and fully studied in \cite{brey2004energy,brey2005scaling}. For \(\alpha=1\), it is \(a_{33}=0\) and no global energy fluctuations exist, as expected.

\subsection{Correlations for the elastic case}
For \(\alpha=1\) the solution given by Eq.~\eqref{eq:smm1q} provides the correlations and fluctuations in the elastic case. It is readily seen that
\begin{equation}
  \label{eq:rhophi0}
\mean{\delta \rho\delta \phi^*}=0.
\end{equation}
Keeping up to order \(k^3\), the other quantities read:
\begin{eqnarray}
  && \frac{\mathcal N}{\mathcal V^2} \mean{\delta \rho\delta \rho^*}\simeq 1-\frac{2\left[(d-1)(d+2)^3\eta^3+8\kappa^3\right]}{(d+2)^2\left[(d-1)(d+2)\eta+2\kappa\right]}k^2, \\
  &&\frac{\mathcal N}{\mathcal V^2} \mean{\delta \rho\delta \epsilon^*}\simeq  1-\frac{2\left[(d-1)(d+2)^3\eta^3+8\kappa^3\right]}{d(d+2)\left[(d-1)(d+2)\eta+2\kappa\right]}k^2, \\
  &&\frac{\mathcal N}{\mathcal V^2} \mean{\delta \phi\delta \phi^*}\simeq  \frac{1}{2}- \nonumber \\ && \quad \frac{(d-1)(d+2)\left[3d(d+2)\eta^3+4\eta^2\kappa\right]+8\kappa^3}{d(d+2)\left[(d-1)(d+2)\eta+2\kappa\right]}k^2, \\
  &&\frac{\mathcal N}{\mathcal V^2} \mean{\delta \phi\delta \epsilon^*}\simeq  \frac{8(d-1)\eta \kappa \left[-(d+2)^2\eta^2+4\kappa^2\right]}{d^2(d+2)\left[(d-1)(d+2)\eta+2\kappa\right]}k^3,  \\
  \label{eq:epep0}
  &&\frac{\mathcal N}{\mathcal V^2} \mean{\delta \epsilon\delta \epsilon^*}\simeq  \frac{d+2}{d} \\ \nonumber  && -\frac{2\left\{(d-1)(d+2)\left[(d+2)^3\eta^3+8\eta \kappa^2\right]+8(d+4)\kappa^3\right\}}{d^2(d+2)\left[(d-1)(d+2)\eta+2\kappa\right]}k^2,
\end{eqnarray}
where the transport coefficients \(\eta\) and \(\kappa\) take their elastic values.

It is important to note that, in the elastic limit, the dependence of the fluctuations and correlations on \(k\) disappear when noise is suppose to be white. This way, Eqs.~\eqref{eq:rhophi0}--\eqref{eq:epep0} provide the first \(k\)-corrections to the classical results \cite{landau2013fluid} when exponential decay of the noise correlations are taking into account. Moreover, the corrections also disappear in the limit \(\eta,\, \kappa \ll 1\) which is related to the conditions \(|\lambda_4|,\, |\lambda_5|\gg 1\), as it is evident from Eqs.~\eqref{eq:lam4} and \eqref{eq:lam5}. 

\subsection{Correlations and fluctuations for \(k>0\) and \(\alpha<1\)}
In the general case, i.e. \(k>0,\) and \(\alpha<1\), the exact expressions for the correlations are given by Eq.~\eqref{eq:smm1q}. Particularly, the density and the longitudinal velocity are also found to be uncorrelated:
\begin{equation}
\label{eq:rhoal}
  \mean{\delta \rho \delta \phi^*}=0.
\end{equation}

For the other magnitudes,  we can not take the limit of small \(k\) directly. Since we are assuming that the HCS is stable, i.e. that \(k^2>\zeta_0/(2\eta)\), we take \(k^2\sim \zeta_0\sim \mu \) small and of the same order, while the other transport coefficients are taken of zeroth order in \(k\). This way, we can show the explicit contribution of the dissipation, at least in its weak limit, as
\begin{eqnarray}
   \frac{\mathcal N}{\mathcal V^2} \mean{\delta \rho\delta \rho^*}&\simeq& 1 + \frac{\frac{d}{4}\zeta_0}{\left(\kappa k^2-\frac{d}{4}\zeta_0\right)}\times \nonumber \\ && \times \frac{\frac{2(d-1)\eta -(d-2)\kappa}{(d-1)(d+2)\eta +2\kappa} k^2 +k_0^2 }{k^2+k_0^2}, \\
  \frac{\mathcal N}{\mathcal V^2}\mean{\delta \rho\delta \epsilon^*} &\simeq& 1 -\frac{k_0^2}{k^2+k_0^2}, \\
  \frac{\mathcal N}{\mathcal V^2}\mean{\delta \phi\delta \phi^*}&\simeq& \frac{1}{2}\left(1 -\frac{k^2_0}{k^2+k_0^2}\right), \\
  \label{eq:phiepk}
  \frac{\mathcal N}{\mathcal V^2} \mean{\delta \phi\delta \epsilon^*}&\simeq&\frac{\frac{4[(d^2-1)\eta+\kappa]}{d^2}k_0^2}{k^2+k_0^2}k,  \\
  \label{eq:epal}  
  \frac{\mathcal N}{\mathcal V^2} \mean{\delta \epsilon\delta \epsilon^*}&\simeq&\frac{d+2}{d}\left(1-\frac{k_0^2}{k^2+k_0^2}\right),
\end{eqnarray}
where 
\begin{equation}
\label{eq:k02}
    k_0^2=\frac{d^2\zeta_0}{4\left[(d-1)(d+2)\eta +2\kappa\right]},
\end{equation}
and \(\eta\) and \(\kappa\) evaluated at \(\alpha=1\). 

It is worth noting that the previous results coincide with that obtained using the approximation of white noise. Moreover, if we set \(\zeta_0=0\) then \(k_0=0\) and we recover the leading orders of the elastic case, Eqs.~\eqref{eq:rhophi0}--\eqref{eq:epep0} with \(k=0\). Nevertheless, the inelastic contribution is relevant, in general, even in the case of the energy-velocity correlations in Eq.~\eqref{eq:phiepk} which, despite showing a small dependence on \(k\), is in the origin of the Casimir-like forces, as we show in the next section. 

\section{\label{sec:casimir} Casimir-like forces}

The results of the previous sections are used to compute the Casimir-like forces on the ``virtual'' walls of the system. In doing so, we first postulate an expression for the pressure tensor up to second order in the fluctuations. Our final goal is to provide an explicit expression for the fluctuation-induced forces for the two-dimension geometry which we compare against event-driven molecular dynamic simulations.

\subsection{Second order pressure tensor}

We assume that the contributions of the fluctuations to the pressure tensor \(P_{ij}\) around the HCS can be obtained by perturbing its Navier-Stokes expression \cite{brey2001hydrodynamic} as\footnote{The prefactor \(2\) of \(2\eta\) appears because of the way the viscosity is written in its dimensionless form.}
\begin{eqnarray}
  \tilde P_{ij}\equiv \frac{P_{ij}}{n_HT_H}&\simeq &(1+\delta\epsilon)\delta_{ij}-2\eta\sqrt{\frac{1+\delta\epsilon}{1+\delta\rho}}\left(\partial_{\ell_i}\frac{\delta \omega_j}{1+\delta\rho} \right. \nonumber \\ &&  \left.+\partial_{\ell_j}\frac{\delta \omega_i}{1+\delta\rho}-\frac{2}{d}\partial_{\boldsymbol{\ell}}\cdot\frac{\delta \boldsymbol \omega}{1+\delta\rho}\delta_{ij}\right).
\end{eqnarray}
The zeroth order in the fluctuations reads \(n_HT_H\delta_{ij}\) which is the nonfluctuating hydrodynamic pressure. The first-order has zero mean and does not contribute to the Casimir force. Hence, we retain the second order:

\begin{eqnarray}
  && \tilde P_{ij}^{(c)}\simeq 2 \eta \left[\partial_{\ell_i}(\delta \omega_j\delta \rho)+\partial_{\ell_j}(\delta \omega_i\delta \rho)-\frac{2}{d}\partial_{\boldsymbol \ell}\cdot(\delta \boldsymbol \omega\delta \rho)\delta_{ij}\right]\nonumber \\  && \qquad  -\eta(\delta\varepsilon-\delta \rho)\left(\partial_{\ell_i}\delta\omega_j+\partial_{\ell_j}\delta \omega_i-\frac{2}{d}\partial_{\boldsymbol \ell}\cdot\delta \boldsymbol \omega \delta_{ij}\right).
\end{eqnarray}
The actual dimensionless second-order pressure tensor \(P_{ij}^{(c)}\) should have an additional noisy term, not directly related with the fluctuating quantities. However, we expect it to have zero mean, as happens with the first-order contribution \cite{brey2011fluctuating}.

Using the fact that \(\delta \rho\) and \(\delta \boldsymbol \omega\) are uncorrelated, see previous section, we have
\begin{equation}
  \mean{\tilde P_{ij}^{(c)}}\simeq -\eta \mean{\delta \epsilon \left(\partial_{\ell_i}\delta\omega_j+\partial_{\ell_j}\delta \omega_i-\frac{2}{d}\partial_{\boldsymbol \ell}\cdot\delta \boldsymbol \omega \delta_{ij}\right)}.
\end{equation}
After some algebra, this expression can be written using the Fourier quantities as
\begin{equation}
  \label{eq:pijc}
  \mean{\tilde P_{ij}^{(c)}}\simeq \frac{2\eta }{\mathcal V^2}\sideset{}{'}\sum_{\boldsymbol k} \left(\frac{k_ik_j}{k^2}-\frac{\delta_{ij}}{d}\right)k \mean{ \delta\phi \delta \epsilon^*}_{k},
\end{equation}
where we have used that \(\mean{\delta \boldsymbol \varphi \delta \epsilon^*}=0\). The prime of the sum stands for its truncation. Here we assume that this is due to the approach to the microscopic scale, namely when (approximately) \(k>\frac{2\pi \lambda}{\sigma}\), with \(\sigma\) the grain diameter. 

Since the sum in \eqref{eq:pijc} includes, for any \(\boldsymbol k\), the vector \(-\boldsymbol k\)  as well, then 
\begin{equation}
  \mean{\tilde P_{ij}^{(c)}}=0, \quad \mbox{for}\quad i\ne j.
\end{equation}
Moreover, it is readily seen that
\begin{equation}
  \sum_{i=1}^d\mean{\tilde P_{ii}^{(c)}}=0.
\end{equation}

The Casimir-like force \(F_i^{(c)}\) on a given wall normal to the \(i\)-direction comes from the contribution of the fluctuation to the pressure tensor: 
\begin{equation}
  F_i^{(c)}=n_HT_H\mean{\tilde P_{ii}^{(c)}}S_i,
\end{equation}
where \(S_i=V/L_i\) is the area (\(d=3\)) or length (\(d=2\)) of the wall. We shall used the dimensionless Casimir-like force per unit area, which can be written as 
\begin{equation}
  f_i^{(c)}=\frac{F_i^{(c)}}{n_HT_H S_i}\simeq \frac{2\eta }{d\mathcal V^2}\sideset{}{'}\sum_{\boldsymbol k} \frac{dk_i^2-k^2}{k} \mean{ \delta\phi \delta \epsilon^*}_{k}.
\end{equation}
By symmetry consideration, when all sides of the system have the same length \(L_1=\dots=L_d\) we have  
\begin{eqnarray}
\nonumber \sideset{}{'}\sum_{\boldsymbol k} \frac{k_x^2}{k} \mean{\delta\phi \delta \epsilon^*}_{k}&=&\sideset{}{'}\sum_{\boldsymbol k} \frac{k_y^2}{k} \mean{\delta\phi \delta \epsilon^*}_{k}\\ &=&\dots=\sideset{}{'}\sum_{\boldsymbol k} \frac{k}{d} \mean{\delta\phi \delta \epsilon^*}_{k}.
\end{eqnarray} 
Hence, our Casimir-like force is zero for a \(d\)-dimension cube.

\subsection{Casimir forces for a rectangular system (\(d=2\))}

For a \(d=2\) system, using Eq.~\eqref{eq:pijc}, the approximate result of Eq.~\eqref{eq:phiepk}, and some other approximations, the dimensionless Casimir force per unit length on the vertical walls can be written as
\begin{eqnarray}
  f_1^{(c)}&\simeq& -\frac{(3 \eta+ \kappa)\eta \zeta_0}{16(2 \eta+ \kappa)}\frac{\mathcal L_1+\mathcal L_2}{\mathcal L_1\mathcal L_2}\frac{\mathcal L_1^2-\mathcal L_2^2}{\mathcal L_1^2+\mathcal L_2^2} \nonumber \\
  \label{eq:f1c}  &=& -\frac{1}{n_H\sigma} \frac{(3 \eta+ \kappa)\eta \zeta_0}{16(2 \eta+ \kappa)}\frac{L_1+L_2}{L_1 L_2}\frac{L_1^2-L_2^2}{ L_1^2+L_2^2},
\end{eqnarray}
and for the horizontal walls \(f_2^{(c)}=-f_1^{(c)}\). The details of the computation are in Appendix \ref{appen:3}.

The force \(f_1^{(c)}\) is clearly negative for \(L_1>L_2\), positive for \(L_1<L_2\), and zero only for \(L_1=L_2\). This means that fluctuations try to push the system towards the square configuration. 

The theoretical prediction of Eq.~\eqref{eq:f1c} is compared against event-driven molecular dynamic simulations in Fig.~\ref{fig:1}. See further details of the simulations in Appendix \ref{appen:4}. The main theoretical prediction are corroborated and a qualitatively good agreement is found.

\begin{figure}[!h]
  \centering
  \includegraphics[width=\linewidth]{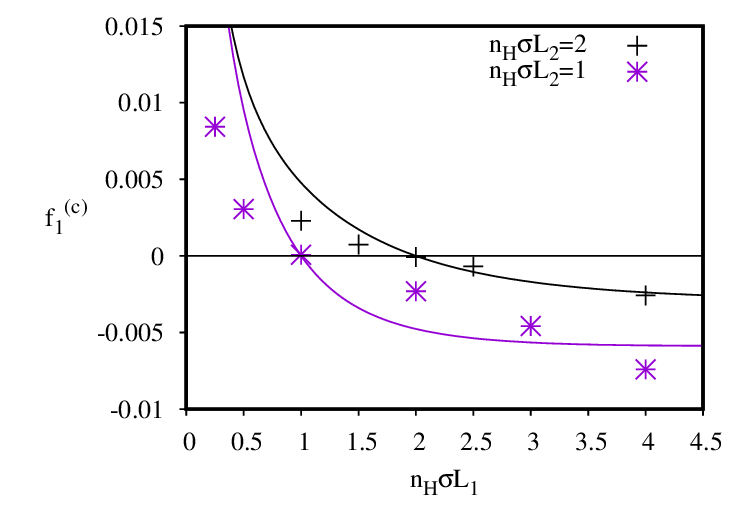}
  \caption{Dimensionless Casimir force per unit length on the vertical wall \(f_1^{(c)}\) as a function of the horizontal length of the system for a fixed density \(n_H\sigma^2=0.025\) and the same coefficient of normal restitution \(\alpha=0.85\). Symbols stands for simulations results and lines for theoretical results of Eq.~\eqref{eq:f1c}. Two values of the height of the system are considered: \(n_H\sigma L_2=1\) (purple asterisks) and \(n_H\sigma L_2=2\) (black crosses).}
  \label{fig:1}
\end{figure}

\section{\label{sec:conclusion}Discussion and conclusions}

In this work, we have studied the long-time limit of the fluctuations and correlations of the density, velocity, and energy around the homogeneous cooling state (HCS), under condition where the HCS is stable. The calculation have been carried out using the fluctuating theory by Brey et. al. \cite{brey2009fluctuating,brey2011fluctuating}, which considers any dimensionality and non-white properties of the noise terms, and is expected to be valid for a dilute gas beyond the limit of small dissipation (\(\alpha\sim 1\)). Our results provide a non-white correction to the classical results in the elastic case [Eqs.~\eqref{eq:rhophi0}--\eqref{eq:epep0}] as well as explicit expressions for small dissipation [Eqs.~\eqref{eq:rhoal}--\eqref{eq:epal}]. Moreover, we have found that non-white contributions can be neglected for small dissipation and small wave vector, provided the HCS keeps stable. 

The dissipation introduces new dependence of the fluctuations and correlations on the wave vector, which vanishes smoothly as we approach the elastic limit. For small dissipation, the main dependence on \(k\) of most correlations comes from the function \(k_0^2/(k^2+k_0^2)\) where \(k_0^2\) is proportional to the cooling rate (dissipation) as given by Eq.~\eqref{eq:k02}. Similar dependence has also been found in driven systems \cite{van1999randomly,brito2007casimir}. However we stress that in our case the previous function is no longer divergent for small \(k\) because, on the one hand,  we can no longer neglect \(k_0\) as it is given by the degree of dissipation and, on the other hand, the stability condition \(k>k_m\simeq k_0\), with \(k_m\) given by Eq.~\eqref{eq:km}, ensures a minimal value of \(k\) which, again, depends on \(\alpha\). This is one more manifestation of the difference between driven and non-driven granular gases, which now relays on the stability of the state around which fluctuations are studied.

The long-time limit of the fluctuations and correlations has been used to compute the fluctuation-induced or Casimir-like force on the walls of the system, by first assuming a form of the pressure tensor up to second order in the fluctuations of the hydrodynamic fields, resulting from perturbing its Naiver-Stokes form. This way, the only contribution to the force comes from the correlations between the longitudinal velocity and the energy, as opposite to what has been proposed in the driven case \cite{Brito_Casimir_Granular,brito2007casimir,shaebani2012nonadditivity} where density-density and density-temperature correlations are the relevant ones. The latter emerges as the perturbation of the hydrostatic pressure (equation of state) as a function of the density and temperature. In our case, a contribution of this kind would produce a systematic deviation of the Casimir-like force, which has not been observed in simulations. The selection of the temperature instead of the energy in the fluctuating hydrodynamic description seems to play an important role at this point. Further investigation of this aspect has to be carried out in order to clarify the relevance of the two complementary mechanisms, that could be done even using recent experimental setups \cite{pitikaris2022granular,yu2020velocity}.

Our theory predicts a zero Casimir-like force for a cube or square system. Moreover, for the two-dimension case, the force on a vertical wall, for instance, is a decreasing function of the horizontal length of the system, being zero only when the length equals the height, which is confirmed with event-driven simulations, even beyond the limit of small dissipation. Hence, the fluctuations have the important effect of stabilizing the square configuration. A similar effect is also expected for the three-dimension case.  
  
\begin{acknowledgments}
This research was supported by Community of Madrid and Rey Juan Carlos University through Young Researchers program in R\&D (Grant No. CCASSE M2737). 
P. R.-L. acknowledges support from Ministerio de Ciencia e Innovaci\'on (Spain), through project NAUTILUS, grant No. PID2020-112936GB-I0 and from AYUDA PUENTE 2022, URJC.

\end{acknowledgments}

\appendix

\section{\label{appen:1}Transport coefficients}
Here we provide explicit expression of the transport coefficients used along the work. All are dimensionless quantities that depend on the dimensionality \(d\) and the coefficient of normal restitution \(\alpha\).

The cooling rate is 
\begin{equation}
  \zeta_0\simeq \frac{\sqrt{2}\pi^{\frac{d-1}{2}}(1-\alpha^2)}{d\Gamma\left(\frac{d}{2}\right)}\left(1+\frac{3}{16}a_2\right)
\end{equation}
where \(a_2\) is related with the kurtosis of the distribution function of the HCS:
\begin{equation}
  a_2\simeq \frac{16(1-\alpha)(1-2\alpha^2)}{9+24d+(8d-41)\alpha +30\alpha^2(1-\alpha)}.
\end{equation}

The long-time limit of the viscosity is
\begin{equation}
  \eta =\left(\frac{8|I|}{1+a_2}-\zeta_0\right)^{-1}, 
\end{equation}
with
\begin{equation}
  I=-\frac{(2d+3-3\alpha)(1+\alpha)\pi^{\frac{d-1}{2}}}{2\sqrt{2}d(d+2)\Gamma\left(\frac{d}{2}\right)}\left(1+\frac{23}{16}a_2\right).
\end{equation}

The long-time limit of the thermal conductivity \(\kappa\) and the diffusive heat conductivity read 
\begin{eqnarray}
  \kappa &=& \frac{(d+2)(1+2a_2)}{2(2|\lambda_5|-\zeta_0)}, \\
  \mu &=& 2 \kappa - \frac{(d+2)(2+a_2)}{4|\lambda_5|},
\end{eqnarray}
where
\begin{eqnarray}
  \lambda_5&=&\frac{4J+(d+2)\zeta_0}{(d+2)a_2}+\frac{3\zeta_0}{2}, \\
  J&=&-\frac{\pi^{\frac{d-1}{2}}(1+\alpha)}{32\sqrt{2}d\Gamma\left(\frac{d}{2}\right)}\left\{16(d+2)(1-\alpha)\right. \nonumber \\ && \qquad +\left. [70+47d-3(34+5d)\alpha a_2]\right\}.
\end{eqnarray}

Other quantities related with the correlations of the noise sources are 
\begin{eqnarray}
  \lambda_4&=&\zeta_0+\frac{4I}{1+a_2}, \\
  a_{33}&=&\frac{d+1}{2d}+\frac{d+2}{4d}a_2+\left[1+d-6d^2\right. \nonumber \\ && \left.-(10-15d+2d^2)\alpha-2(2+7d)\alpha^2\right. \nonumber \\ && \left.+2(10-d)\alpha^3\right]\left[6d(2d+1)-2d(11-2d)\alpha \right. \nonumber \\ && \qquad \left. +12d\alpha^2-12d\alpha^3 \right]^{-1}.
\end{eqnarray}
and
\begin{eqnarray}
  \label{eq:cw}
  c_w&=&\frac{2(d-1)}{d}\frac{\mathcal V^2}{N}\frac{1+a_2}{4}k^2, \\
  \label{eq:cep1}
  c_\epsilon^{(1)}&=&\frac{4\mathcal V^2}{N}\zeta_0a_{33}, \\
  \label{eq:cep2}
  c_\epsilon^{(2)}&=&\frac{(d+2)\mathcal V^2}{d^2N}\left[1+\frac{d+8}{2}a_2\right. \nonumber \\ && \left.+\frac{2d\zeta_0a_{33}(1+2a_2)}{|\lambda_5|-\frac{\zeta_0}{2}}\right]k^2, 
\end{eqnarray}

\section{\label{appen:2} Computation of the structure factor}

The starting point to compute the correlations among fluctuating fields at the same time and opposite wave vectors, is to define the matrix structure factor as
\begin{eqnarray}
  S(s,k)&=&\mean{\delta \boldsymbol u(s,k)[\delta \boldsymbol u(s,-k)]^t}\nonumber \\ &=&\mean{\delta \boldsymbol u(s,k)[\delta \boldsymbol u(s,k)]^\dag},
\end{eqnarray}
where last equality holds from Eq.~\eqref{eq:realrho}. Taking the time derivative of the structure factor and the formal solution in Eq.~\eqref{eq:deltau} we arrive at the following equation
\begin{equation}
  \label{eq:eqstr}
  \partial_sS=AS+SA^t+\mean{\boldsymbol b \delta \boldsymbol u^\dag+\delta \boldsymbol u \boldsymbol b^\dag},
\end{equation}
where the matrix \(A\) and vector \(\boldsymbol b\) are given by Eqs.~\eqref{eq:matA} and \eqref{eq:vecb}, respectively. The bracket term of the equation can be computed using Eq.~\eqref{eq:deltau}:
\begin{eqnarray}
  \mean{\boldsymbol b \delta \boldsymbol u^\dag+\delta \boldsymbol u \boldsymbol b^\dag}&=& \int_0^s ds_1 \, B(s-s_1) e^{(s-s_1)A^t} \nonumber \\ &&  +\int_0^s ds_1 e^{(s-s_1)A} \, B(s-s_1),
\end{eqnarray}
where we have introduced the real and symmetric correlation matrix
\begin{equation}
  B(s)=\mean{\boldsymbol b(s,k) \boldsymbol b(0,k)^\dag}
\end{equation}
whose dependence on \(k\) has been omitted for simplicity. Using the correlation properties of the noise terms given by Eqs.~\eqref{eq:corrw} and \eqref{eq:correp}, \(B\) can be written as
\begin{eqnarray}
  \label{eq:bs}
  B(s)&\simeq&c_we^{\lambda_4 s}I_2+[c_{\epsilon}^{(1)}\delta(s)+c_{\epsilon}^{(2)}e^{\lambda_5 s}]I_3\\ & \underset{|\lambda_4|,|\lambda_5|\gg 1}{\longrightarrow}& \left[\frac{2c_w}{|\lambda_4|} I_2+\left(c_{\epsilon}^{(1)}+\frac{2c_{\epsilon}^{(2)}}{|\lambda_5|}\right) I_3\right]\delta(s),
\end{eqnarray}
where \(c_w\), \(c_\epsilon^{(1)}\), and \(c_\epsilon^{(2)}\) are given in Eqs.~\eqref{eq:cw}--\eqref{eq:cep2} and the new matrices are 
\begin{equation}
  I_2=
  \begin{pmatrix}
    0 & 0 &0 \\
    0 & 1 &0 \\
    0 & 0 & 0 
  \end{pmatrix}
  ; \qquad
  I_3=
  \begin{pmatrix}
    0 & 0 &0 \\
    0 & 0 &0 \\
    0 & 0 &1 
  \end{pmatrix}
  .
\end{equation}

The steady-state solution to Eq.~\eqref{eq:eqstr} is obtained by setting \(\partial_sS=0\) and taking \(s\to \infty\). By making the change of variable \(\tau=s-s_1\), the steady-state structure factor is given by
\begin{equation}
  AS+SA^t+\int_0^\infty d\tau\,\left[e^{\tau A}B(\tau)+B(\tau)e^{\tau A^t}\right]=0.
\end{equation}
The integral can be carried out by using expression \eqref{eq:bs} of \(B\) and the well-known results
\begin{eqnarray}
  && \int_0^\infty d\tau \, \delta(\tau) f(\tau)=\frac{1}{2}f(0), \\
  &&  \int_0^\infty d\tau\, e^{\tau (\lambda I+A)}=-(\lambda I+A)^{-1},
\end{eqnarray}
valid provided \(\lambda I+A\) has negative spectrum, with \(I\) being the identity matrix, and for any regularly enough function \(f(\tau)\). The final equation for the structure factor \(S\) read
\begin{equation}
  \label{eq:assa}
  AS+SA^t-Q=0,
\end{equation}
with
\begin{eqnarray}
  \nonumber Q=&&c_w\left[I_2(\lambda_4I+A^t)^{-1}+(\lambda_4I+A)^{-1}I_2\right]-c_\epsilon^{(1)}I_3\\  && +c_{\epsilon}^{(2)}\left[I_3(\lambda_5I+A^t)^{-1}+(\lambda_5I+A)^{-1}I_3\right] \\ &&\underset{|\lambda_4|,|\lambda_5|\gg 1}{\longrightarrow} -\frac{2c_w}{|\lambda_4|} I_2-\left(c_\epsilon^{(1)}+\frac{2c_\epsilon^{(2)}}{|\lambda_5|}\right)I_3.
\end{eqnarray}

Thank to the symmetric structure of the matrix equation \eqref{eq:assa} for the structure matrix \(S\), it can be written in a more familiar form by using the relevant entries of \(S\) to construct the structure vector \(\boldsymbol S\) given by Eq.~\eqref{eq:sfv}:
\begin{equation}
  \label{eq:msq}
  M\boldsymbol S=\boldsymbol Q,
\end{equation}
where \(M\) and \(\boldsymbol Q\) are given in Eqs.~\eqref{eq:matm} and \eqref{eq:matq}, respectively. When \(M\) is not singular, the solution to Eq.~\eqref{eq:msq} is given by Eq.~\eqref{eq:smm1q}.

\section{\label{appen:3} Approximate computation of the Casimir forces}

Using Eq.~\eqref{eq:pijc} with Eq.~\eqref{eq:f1c} and the result of Eq.~\eqref{eq:phiepk}, valid for small dissipation and small values of \(k\), the dimensionless Casimir force per unit length for a \(d=2\) system reads
\begin{equation}
  f_1^{(c)}\simeq \frac{(3 \eta+ \kappa)\eta \zeta_0}{2(2 \eta+ \kappa) N} \sideset{}{'}\sum_{\boldsymbol k}\frac{k_1^2-k_2^2}{k^2+k_0^2},
\end{equation}
where
\begin{equation}
  k_0^2=\frac{\zeta_0}{2(2 \eta+ \kappa)},
\end{equation}
and the prime indicates a truncation of the sum, as specified bellow. The sum can be written as 
\begin{equation}
  \sideset{}{'}\sum_{\boldsymbol k}\frac{k_1^2-k_y^2}{k^2+k_0^2}=\sideset{}{'}\sum_{n_1,n_2}S_{n_1,n_2}
\end{equation}
with
\begin{equation}
  S_{n_1,n_2}=\frac{\left(\frac{2\pi}{\mathcal L_1}n_1\right)^2-\left(\frac{2\pi}{\mathcal L_y}n_2\right)^2}{\left[\left(\frac{2\pi}{\mathcal L_1}n_1\right)^2+\left(\frac{2\pi}{\mathcal L_y}n_2\right)^2\right]+k_0^2}.
\end{equation}
Taking into account that \(n_1\) and \(n_2\) take positive and negative values, we have 
\begin{eqnarray}
  \sideset{}{'}\sum_{\boldsymbol k}\frac{k_1^2-k_2^2}{k^2+k_0^2}&=&\sideset{}{'}\sum_{n_1,n_2}S_{n_1,n_2}\simeq \frac{1}{2}\sideset{}{'}\sum_{n_1,n_2}\left(S_{n_1,n_2}+S_{n_2,n_1}\right)\nonumber \\ && = 2\sideset{}{'}\sum_{n_1>0,n_2>0}\left(S_{n_1,n_2}+S_{n_2,n_1}\right) \nonumber \\ && \simeq 2 \int_0^{n_M} dn_1 dn_2\,\left(S_{n_1,n_2}+S_{n_2,n_1}\right).
\end{eqnarray}
In the first approximation of the previous equation we assume that the truncation of the sum is as \(n_1, \,n_2<n_M\) and not over the values of \(k_1\) and \(k_2\), with
\begin{equation}
  n_M \simeq \frac{\mathcal L_1+ \mathcal L_2}{4\pi n_H\sigma^2}.
\end{equation}
The second approximation is an integral approximation of the sum. In order to compute the integral, we introduce further approximations:
\begin{eqnarray}
  && S_{n_1,n_2}+S_{n_2,n_1}=\left[\left(\frac{2\pi}{\mathcal L_1}\right)^2-\left(\frac{2\pi}{\mathcal L_2}\right)^2\right] \nonumber \\ && \quad \times \frac{2\left[\left(\frac{2\pi}{\mathcal L_1}\right)^2+\left(\frac{2\pi}{\mathcal L_2}\right)^2\right]n_1^2n_2^2+k_0^2(n_1^2+n_2^2)}{\left[\left(\frac{2\pi n_1}{\mathcal L_1}\right)^2+\left(\frac{2\pi n_2}{\mathcal L_2}\right)^2+k_0^2\right] \left[\left(\frac{2\pi n_2}{\mathcal L_1}\right)^2+\left(\frac{2\pi n_1}{\mathcal L_2}\right)^2+k_0^2\right]} \nonumber \\ && \quad \simeq 2 \frac{\mathcal L_2^2-\mathcal L_1^2}{\mathcal L_1^2+\mathcal L_2^2}\frac{4n_1^2n_2^2+\frac{2k_0^2}{\left(\frac{2\pi}{\mathcal L_1}\right)^2+\left(\frac{2\pi}{\mathcal L_2}\right)^2}(n_1^2+n_2^2)}{\left[n_1^2+n_2^2+\frac{2k_0^2}{\left(\frac{2\pi}{\mathcal L_1}\right)^2+\left(\frac{2\pi}{\mathcal L_2}\right)^2}\right]^2},
\end{eqnarray}
which is valid provided \(\mathcal L_1^2\simeq \mathcal L_2^2\simeq \mathcal L_1\mathcal L_2\). Finally, assuming that \(\frac{2k_0^2}{\left(\frac{2\pi}{\mathcal L_1}\right)^2+\left(\frac{2\pi}{\mathcal L_2}\right)^2}\sim 1\) and \(n_m\gg 1\) and taking into account that
\begin{equation}
  N=n_H L_1 L_2=\frac{1}{n_H\sigma^2}\mathcal L_1\mathcal L_2,
\end{equation}
we arrive at Eq.~\eqref{eq:f1c}.

\section{\label{appen:4} Event-driven numerical simulations}

We have run numerical simulation using the event-driven algorithm by Lubachevsky \cite{lubachevsky1991simulate}, adapted to account for the inelastic collisions and the steady-state representation \cite{lutsko2001model,brey2004steady}. All simulations started with an initial condition where all particles are homogeneously distributed and have Gaussian velocity distribution. After a transient, of the order of \(10^4\) collisions per particle, measurements are done along \(10^5\) collisions per particle. In addition, quantities are obtained after averaging over different times as well as different realizations.

The forces on the ``virtual'' walls have been measured as the average momentum transfer of particles. Results are shown in the following figure for \(\alpha=1\) (elastic case) and \(\alpha=0.85\) and different system sizes. It is observed that in the elastic case there is no dependence on the system size (no Casimir-like force), within the numerical precision. In the inelastic case, the is no dependence on the system size, provided the system remains square \(L_1=L_2\), while the horizontal force and the vertical one are symmetric with respect to their hydrodynamic values with rectangular geometries. This allowed as to identify the Casimir-like forces \(f_1^{(c)}\) and \(f_2^{(c)}\), by subtracting the force of square configurations from the total force.

\begin{figure}[!h]
  \centering
  \includegraphics[width=\linewidth]{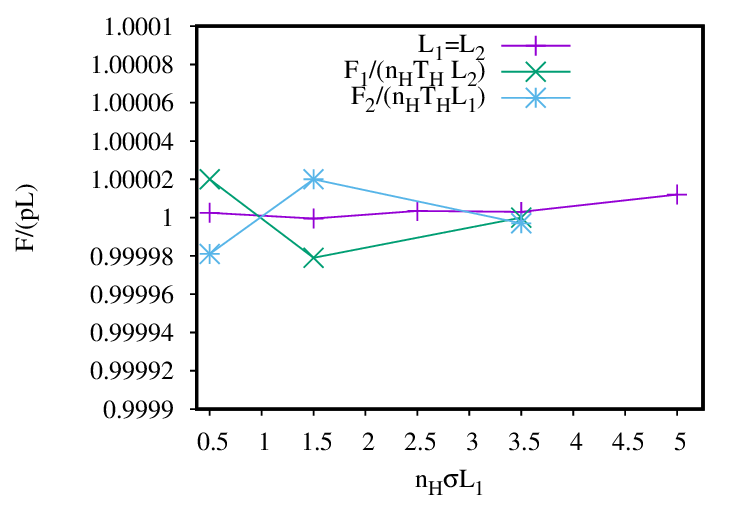}
  \includegraphics[width=\linewidth]{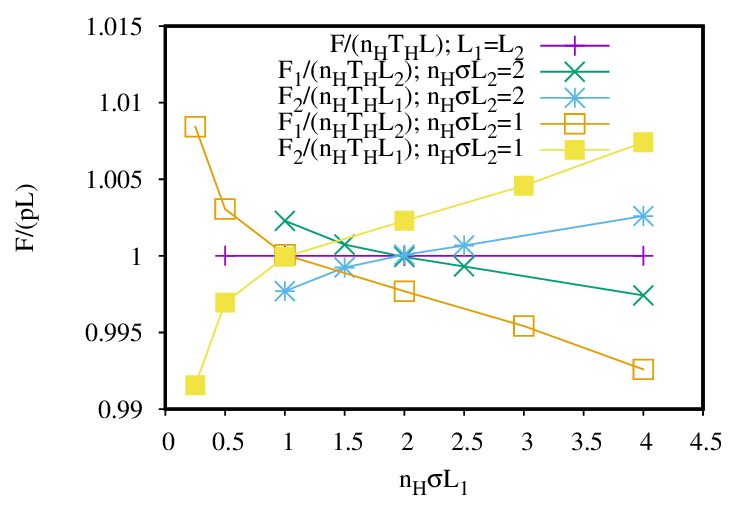}
  \caption{Numerical simulations results for the total (nonfluctuating and fluctuating) force on the vertical walls for the elastic case \(\alpha=1\) (top) and \(\alpha=0.85\) (bottom) with \(n_H\sigma^2=0.025\).}
  \label{fig:2}
\end{figure}

\nocite{*}
\bibliography{biblio}

\end{document}